\documentclass[aps,preprint,showpacs]{revtex4-1}
\usepackage{ifpdf}
\usepackage{pdfsync}
\ifpdf
\usepackage{hyperref}
\else
\usepackage[hypertex]{hyperref}
\fi
\usepackage{graphicx,subfigure}
\usepackage{xcolor}
\usepackage{amsmath,amsfonts,amssymb}

\newcommand{\f}{\frac}

\newcommand{\intl}{\int\limits}

\graphicspath{{D:/localtexmf/Mytex/LGH/mathplots/},{D:/localtexmf/Mytex/LGH/eqstcan/zeno_jcp/zeno_timmerman/}}
\begin{document}
\title{The critical compressibility factor value: associative fluids and liquid alkali metals}
\author{V.L. Kulinskii}
\email{kulinskij@onu.edu.ua}
\affiliation{Department for Theoretical
Physics, Odessa National University, Dvoryanskaya 2,
65026 Odessa, Ukraine}
\begin{abstract}
We show how to obtain the critical compressibility factor $Z_c$
for simple and associative Lennard-Jones fluids using the
critical characteristics of the Ising model on different lattices. The results show that low values of critical compressibility factor are correlated with the associative properties of fluids in critical region and can be obtained on the basis of the results for the Ising model on lattices with more than one atom per cell. An explanation for the results on the critical point line of the Lennard-Jones fluids and liquid metals is proposed within the global isomorphism approach.
\end{abstract}
 \maketitle

\section{Introduction}\label{sec_intro}
The classical Principle of Corresponding States (PCS)
is widely used to study the thermodynamic properties
of matter \cite{pcs_guggenheim_jcp1945,book_filippovthermophys_1988en}. In its general statement the PCS can be proven explicitly
only if the interaction potentials of the systems can
be related by simple scaling transformation of the
parameters \cite{book_prigozhisolut}. For real
systems the interatomic potential is very complicated
due to the complex structure of molecules.
Moreover, the effective interaction varies with the
density and temperature due to excitation of the
electronic degrees of freedom. This situation is
typical for liquid metals \cite{book_liqmetals_march}.
In such cases the refinement of the na\"{\i}ve form
of the PCS is needed \cite{book_filippovthermophys_1988en,eos_riedelpcs1_cingtech1954,liq_pcspitzer2_jamchemsoc1955}.
In comparison with the simple molecular
fluids the behavior of associative fluids and liquid
metals is much more diverse. These systems are
interesting in view of the deviation from the
classical formulation of the PCS based on simple
scaling of the interaction potential. In particular, the coexistence curves (CC) of alkali metals like $Cs, Rb$ has greater asymmetry \cite{liqmetals_hensel_jpcondmat1990,crit_coulomb_pitzer_jpc1995}. Nevertheless, the general properties of the binodal appear to be the same as those of simple fluids. Namely, the binodal shows the same global approximate cubic behavior and obeys the law of rectilinear diameter \cite{crit_diam0} (LRD) in wide temperature range below $T_c$:
\begin{equation}\label{rdl}
  n_{d} = \f{n_{l}+n_{g}}{2\,n_c} = 1+A\,
\left(\, 1-T/T_c \,\right)\,.
\end{equation}
Here $n_{l,g}$ is the number density of the liquid and vapor phases correspondingly, $n_{c}$ is the critical density.

In addition to the LRD there is one more remarkable phenomenological linearity - the linear character of the unit compressibility factor line:
\begin{equation}\label{z1}
Z = \f{P}{n\,T}\,\,\Rightarrow\,\,  \f{T}{T_{Z}}+\f{n}{n_Z} = 1\,.
\end{equation}
which is called the Zeno-line (``Zeno`` stands for ``$Z$ equals one``  \cite{eos_zenobenamotz_isrchemphysj1990}).
It can be derived easily for the van der Waals equation of state and in this case known as the Batschinski law \cite{eos_zenobatschinski_annphys1906}.
Here $T_Z$ and $n_Z$ are determined as follows \cite{eos_zenobenamotz_isrchemphysj1990}:
\begin{equation}\label{tbnb}
  B_2(T_{Z}) = 0\,,\quad n_Z= \f{ T_Z }{B_3\left(\,T_Z\,\right)}\,\left. \f{dB_2}{dT}\right|_{T= T_Z}\,.
\end{equation}
Clearly $T_Z$ is the Boyle temperature.

For simple fluids and model systems with pair interactions of the Lennard-Jones (LJ) type and even for such associative liquids like ammonia and water Zeno-linear law holds fairy well as the available data show \cite{eos_zeno_jphyschem1992,eos_zenoboyle_jphysc1983,*water_zenoline_ijthermophys2001}.
Thus linearities Eq.~\eqref{rdl} and Eq.~\eqref{z1} can serve as a basis for the extended formulation of the PCS with thermodynamic similarity classes determined by these linear laws. The validity of these linearities for broad set of liquid metals and molecular fluids was studied in a series of works of Apfelbaum and Vorob'ev \cite{eos_zenoapfelbaum_jpchemb2006,eos_zenoapfelbaum1_jpcb2009,eos_zenoline_potentials_jcp2009}.
In particular in \cite{liqmetals_zenoapfelbaum2_cpl2009,eos_vorobevglobaliso_jpc2012,*eos_apfelberill_jpcb2012,*thermodyn_morse_jcp2011} it was demonstrated that liquid metals also obey extended thermodynamic similarity based on the Zeno-line linearity. The correlation of the Zeno-parameters $T_Z, n_Z$ with the critical parameters $T_c, n_c$ similar to that in molecular fluids was demonstrated. Using such correlation the position of high-temperature critical points inaccessible via direct measurements can be estimated for metals like $Al, Cu, Fe$ etc. These estimates was based on the results \cite{eos_zenoapfelbaum_jpchemb2008,eos_zenoline_potentials_jcp2009} which state the relation between the critical point parameters and those of the Zeno-line \cite{eos_zenobenamotz_isrchemphysj1990}. According to the general idea of similarity, in a sense of Apfelbaum\&Vorob'ev approach, the ratios $T_c/T_Z$, $n_c/n_Z$ should be universal.

In \cite{eos_zenome_jphyschemb2010} it was shown that the results of Apfelbaum\&Vorob'ev can be interpreted as the correspondence between the phase diagram of the fluid and the lattice gas of the form:
\begin{equation}\label{projtransfr_my}
  n =\, n_*\,\f{x}{1+z \,\tilde{t}}\,,\quad
  T =\, T_*\,\f{z\, \tilde{t}}{1+z \,\tilde{t}}\,,
\end{equation}
The key element in constructing transformation Eq.~\eqref{projtransfr_my} is the linear Zeno-element
\begin{equation}\label{eq_zenoelement}
  \f{n}{n_*}+  \f{T}{T_*} = 1
\end{equation}
which is associated with the line $x=1$ of the lattice gas model. Here $x$ is the lattice gas density, $\tilde{t} = t/t_c$ is the dimensionless temperature of the lattice model so that the critical state is given by $\tilde{t}_c = 1$ and $x_c = 1/2$. The parameter $z$ is
\begin{equation}\label{ztc}
    z = \f{T_c}{T_*-T_c}\,,
\end{equation}
and can be related with the scaling property of the
attractive potential
$\Phi_{\text{attr}}(r)\propto -1/r^{m}$: $z = d/m$,
where $d$ is the dimension \cite{eos_zenomeglobal_jcp2010}.
The temperature parameter $T_*$ is the Boyle temperature in the vdW approximation \cite{book_ll5_en}:
\begin{equation}\label{tb_vdw}
  T_{*} = \f{a}{b}\,,
\end{equation}
where
\begin{align}\label{vdw_ab}
a =\,\, -2\pi\,\intl_{\sigma}^{+\infty}\Phi_{attr}(r)\,r^2\,dr\,.
\end{align}
and $b= \f{2\pi}{3}\,\sigma^{3}$, $\sigma$ is the effective diameter of the particle hard core. The density parameter $n_*$
is determined by the following relation:
\begin{equation}\label{nbme}
n_*= \f{ T_* }{B_3\left(\,T_*\,\right)}\,\left. \f{dB_2}{dT}\right|_{T= T_*}\,.
\end{equation}
The Zeno-element parameters $n_*$ and $T_*$
are different from $n_{Z}, T_{Z}$ because linear
Zeno-element do not coincide with the Zeno-line. Only in case of vdW EoS they become equal. E.g. for the LJ potential
$V_{LJ}(r) = -4\,\varepsilon\,
\left(\, (r/\sigma)^{-6} - (r/\sigma)^{-12} \,\right)$ in
3D:
\begin{equation}\label{nbtb3d}
T_*/\varepsilon = 4\,\,,\quad n_*\,\sigma^3 \approx 0.976
\end{equation}
and $z=1/2$ \cite{eos_zenomeglobal_jcp2010}. Further we will use the dimensionless variables $n \to n\,\sigma^{3}\,,\,\,T\to T/\varepsilon$. Basing on such simple geometrical scheme the estimates for the CP loci of the Lennard-Jones (LJ) fluids were obtained in different dimensions \cite{eos_zenomeunified_jphyschemb2011,eos_vliegerthartlekkerkerkerme_jcp2011}.
Indeed, the mapping \eqref{projtransfr_my} gives 1-1 correspondence between liquid-gas states of a fluid and the lattice gas model, so the critical point $x_c = 1/2, \tilde{t}_c = 1$ of the lattice gas transforms to the CP of the LJ fluid:
\begin{equation}\label{nctcmy}
  n_{c} = \f{n_*}{2\left(\,1+z\,\right)}\,,\quad T_{c} = T_*\, \f{z}{1+z}\,.
\end{equation}
E.g. for 3D case Eq.~\eqref{nctcmy} leads to the estimates:
\begin{equation}\label{nctcmy3d}
  n_{c} = n_{*}/3 \approx 0.32\,,\quad T_{c} = T_*/3 \approx 1.33\,,
\end{equation}
which correlate quite good (the difference less than 5\% ) with known numerical results \cite{crit_lj3dim_jcp1998}. It should be noted that the approach on which Eq.~\eqref{projtransfr_my} is based does not include the critical fluctuations. It is mean-field like in physical sense as it does not use the renormgroup
symmetry of the critical fluctuations \cite{crit_wilsonkogut_prep1974}.

Recently, it was demonstrated that using Eq.~\eqref{projtransfr_my} it is possible to obtain the relation between the critical compressibility factors $Z_c$ of the LJ-fluid and the lattice gas, or equivalently the Ising model \cite{eos_zenozcme_jcp2013}.

The aim of this paper is to extend the result of \cite{eos_zenozcme_jcp2013} obtained for simple fluids
to liquid alkali metals and other fluids like water, ammonia and methanol which demonstrate the associative properties in near critical region. We propose an explanation for the difference between the values of the critical compressibility factor $Z_c$ of simple fluids and that of liquid metals, water, ammonia and methanol.  The idea is  that it is caused by the difference between the lattice structures of the isomorphic lattice models, namely, the number of particles per unit cell that reflects the associative properties of a fluid. Also we derive the equation for the critical point line of the Lennard-Jones (LJ) fluids consistently from the global isomorphism approach.

The paper is organized as follows. In Section~II we discuss the isomorphism transformation and its parameters. We show that the use of corresponding parameters of the projective mapping  unifies the description of CP parameters for simple LJ-fluids and liquid metals. We propose the explanation for the difference in the critical compressibility factor values of simple liquids and associative liquids like liquid metals, water etc. using global isomorphism approach. In Section~\ref{sec_weakstrict} we derive the equation of the line of the critical points for the LJ fluids.  and give the consistent interpretation of the results in \cite{eos_zenoapfelbaum1_jpcb2009}.

\section{Critical compressibility factor for fluids and liquid metals}\label{sec_zc}
The critical compressibility factor
$Z_c = P_c/(n_c\,T_c)$ is the basic parameter
for the definition of thermodynamic similarity class.
It is invariant with respect to the scaling transformation
of the parameters of the interaction.
The substances are commonly classified with respect
to this value. As is known for simple fluids $Z_c = 0.28\div 0.30$ and differs substantially from that of the vdW EoS
$Z_c = 0.375$.

In \cite{eos_zenozcme_jcp2013} the relation between critical compressibility factor (CCF) of LJ-fluid and the isomorphic lattice model (ILM) was derived:
\begin{equation}\label{zmy}
  Z^{(fl)}_{c} = \f{P_c}{n_c\,T_{c}} =\f{(1+z)^2}{z}\,\f{t_c}{T_*}\, Z^{(LG)}_{c}\,\,.
\end{equation}
Here $t_c$ is the critical temperature of the Ising model and $Z^{(LG)}_{c}$ is the compressibility factor of the lattice gas model. Eq.~\eqref{zmy} is applicable at least for the fluids where the thermodynamics is governed mainly by the van der Waals dispersive interaction with $r^{-6}$ asymptote, in particular, for the LJ fluids (see \cite{eos_zenogenpcs_jcp2010}).

The formula \eqref{zmy} is obtained from the following relation between the thermodynamic potentials of the LJ fluid and the
Lattice Gas:
\begin{equation}\label{potpot}
  P(T,\mu)\,V = \mathfrak{G}(t(T),h(\mu,T))\,,
\end{equation}
where $P$ is the pressure of a fluid, $\mu$ - its chemical potential, $\mathfrak{G}$
is the Gibbs potential of the lattice model, $h$ is the field conjugated to the order parameter $x$. This is direct
consequence of Eq.~\eqref{projtransfr_my} \cite{eos_zenomeunified_jphyschemb2011}.

In accordance with \cite{book_rice_thermodyn} the value $Z^{(LG)}_{c}$ is related with the partition function per spin of the critical state $G^{1/N}_{c}$ of the Ising model:
\begin{equation}\label{eq_zpartition}
Z^{(LG)}_{c} = 2\,\ln G_{c}^{1/N}
\end{equation}
E.g. in 3D Ising model case of cubic lattice with isotropic interaction: $Z^{(LG)}_{c} = 0.221$, the critical temperature is $t_c \approx 4.51\,J$ \cite{crit_3disingliufisher_physa1989}. If we take into account that the interaction constants of the LJ potential $\varepsilon$ and the Ising model $J$ connected by the relation $\varepsilon = 4\,J$ then Eq.~\eqref{zmy} along with Eq.~\eqref{nbtb3d} gives:
\begin{equation}\label{zmy_lj3d}
  Z^{(fl)}_{c}\approx 1.27\, Z^{(LG)}_{c} = 0.28\,,
\end{equation}
which correlates quite well with the CCF value for
simple fluids like noble gases (see Table~\ref{tab_zc_simple}).
\begin{table}
  \centering
  \begin{tabular}{|c|c|c|c|c|}
    \hline
Fluid &
\hspace{0.2cm} $Ne$ \hspace{0.2cm}&
\hspace{0.2cm} $Ar$ \hspace{0.2cm}&
\hspace{0.2cm} $Kr$ \hspace{0.2cm}&
\hspace{0.2cm} $Xe$ \hspace{0.2cm}\\
\hline
$Z_c$ &0.30&0.29&0.29&0.29\\
\hline
  \end{tabular}
  \caption{Experimental values of $Z_c$ for simple  fluids \cite{book_filippovthermophys_1988en,book_vargaftik1975}.}\label{tab_zc_simple}
\end{table}

The difference in values of $Z_c$ for the noble gases and other complex liquids (see Table~\ref{tab_zc_cmplx}) is usually attributed to more complex interactions in liquids. Such fluids do not belong to the similarity class of the noble fluids.
Empirical correcting parameters like the Pitzer's acentric factor $\omega$ or Riedel's factor are introduced to characterize this difference \cite{liq_pcspitzer2_jamchemsoc1955,eos_riedelpcs1_cingtech1954}. E.g. the Pitzer's acentric factor $\omega$ reflects the nonspherical shape of the particles though its definition is of pure macroscopical nature in thermodynamic terms.
The results of \cite{eos_zenozcme_jcp2013} allow to get the dependence of $Z_{c}$ on the Pitzer's acentric factor $\omega$ which reproduces the data for the liquids quite well for $\omega < 0.3$. Greater values of $\omega$, which lead to the lower values of $Z_c$, obviously cannot be interpreted directly in microscopic terms because of the phenomenological nature of this parameter. So the usage of the Pitzer's acentric factor for the fluids with low values of $Z_{c}$ (see Table~\ref{tab_zc_cmplx}) masks the physical mechanism behind deviation from the noble fluids value $Z_{c} \approx 0.29$. It should be noted that low compressibility factor values for water, methanol as well as liquid alkali metals
correlate with their associative properties.
The molecular dimers and higher associates are formed in the near critical region of these fluids \cite{liqmetals_assoc_jphyschem1967,dimers_methanol_jpc1971,liqmetals_hensel_jpcm1988,water_dimer_us_nato2007}.
\begin{table}
\center
\begin{tabular}{|c|c|c|c|c|c|c|c|c|c|}
\hline
& $Li$ & $Na$ & $K$ & $Rb$ & $Cs$ & $Hg$, \cite{liqmetals_henzel_zphyschem1988}&$CH_3OH$, \cite{book_nist69}&$H_2O$, \cite{book_nist69}&$NH_3$, \cite{book_nist69} \\
  \hline
$Z_c$& 0.21 & 0.23 & 0.21 & 0.22 & 0.21 & 0.39 & 0.19 & 0.23 & 0.25\\
\hline
\end{tabular}
\caption{Experimental values of $Z_c$ for some molecular fluids and liquid metals \cite{book_vargaftik1975}.}\label{tab_zc_cmplx}
\end{table}

From the point of view of the global isomorphism
approach it is natural to relate the associative
property of a
fluid and the structure of the lattice.
We put forward the simplest assumption that the association of a fluid can be interpreted in terms of the isomorphic lattice with more than 1 particle per unit cell taking into account the association factor $q$ - the number of primitive particles in associates. The prevalence of dimers means that
$q=2$, trimers $q=3$ etc. Obvious options for associative
fluids are the simple cubic,
the body centered cubic and the
face centered cubic with $q=1,2,4$ correspondingly.

Taking into account that for the associative fluid
the interactive unit consists of $q$ ``primitive``
particles Eq.~\eqref{zmy} is changed into:
\begin{equation}\label{zmy_c}
  Z^{(fl)}_{c} = \f{1}{q}\,\f{(1+z)^2}{z}\,\f{t_c}{T_*}\, Z^{(LG)}_{c}\,\,.
\end{equation}
The results for the lattices are given in
Table~\ref{tab_zclatt}.
\begin{table}
\center
\begin{tabular}{|c|c|c|c|c|}
\hline
    &\hspace{0.2cm} $q$ \hspace{0.2cm}& \hspace{0.2cm} $t_c$ \hspace{0.2cm}& \hspace{0.2cm} $Z^{(LG)}_c$ \hspace{0.2cm} & \hspace{0.2cm}  $Z^{(fl)}_{c}$ \hspace{0.2cm}\\
\hline
\hline
sc  & 1 & 4.51 & 0.221 & 0.281\\
\hline
bcc & 2 & 6.35 & 0.239 & 0.214\\
\hline
fcc & 4 & 9.80 & 0.246 & 0.17\\
\hline
\end{tabular}
\caption{Critical parameters ($t_c$ and $Z^{(LG)}_c$) of the isotropic Ising model on different 3D lattices taken from \cite{crit_dombsykes_prc1956,crit_zclattice_ptp1989} and corresponding values of $Z^{(fl)}_c$ for the LJ fluid obtained using Eq.~\eqref{zmy_c}.}\label{tab_zclatt}
\end{table}
The comparison of theoretical estimates Eq.~\eqref{zmy_c} for $Z_c$ shows good correlation with the data (see Table~\ref{tab_zc_cmplx}).

It should be noted that for light liquid metals with high temperature CP there is uncertainty in value $Z_c$ because direct measurements are not possible. The extrapolation methods for the low temperature data are used to obtain the parameters of the critical state. For example, for lithium $Z_c \approx 0.17$, for sodium $Z_{c}\approx 0.14$ according to \cite{liqmetals_bystrovshpilrain_1990}. Commonly, such extrapolation exploits the linear diameter law \eqref{rdl}(see e.g. \cite{book_frenkelsimul}).
In fluctuational region the classical, mean-field behavior of the diameter breaks down and non analytical terms $\propto |\tau|^{2\beta}$ and $\propto |\tau|^{1-\alpha}$ ($\tau = (T-T_c)/T_c$) should be taken into account \cite{crit_rehrmermin_pra1973,crit_can_kul_cmphukr1997,crit_yydiamfisherorkoulas_prl2000}.

The asymmetry of the binodal for alkali metals is bigger than that of molecular fluids and it appears that the amplitudes of such non analytical terms are also big in comparison with those of molecular fluids \cite{liqmetals_singdiamhensel_prl1985}.
The symmetrization of the binodal with the nonlinear fluctuational terms like $|\tau|^{1-\alpha}$ and $|\tau|^{2\beta}$ should be distinguished from
the regular nonlinearity and needs different approach (see e.g. \cite{crit_can_diamsing_kulimalo_physa2009}).
In principle, such terms can be accounted for within the M.~Fisher's theory of complete scaling \cite{crit_fisherdiam_chemphyslet2005} (see also \cite{crit_diam_schroerweiss_statphys2008} and reference therein). For high temperature metals like $Zn$, $Pb$ and $Au$ such procedure has been performed recently in \cite{liqmetals_schroer_hthp2014}. The results assume that for $Au$ $Z_c\approx 0.22$, which corresponds to the alkali metal value. The value of CCF for $Zn$ and $Pb$ is $Z_c \approx 0.29$ and therefore comparable with that of the noble gases. In view of the electronic structure it could be explained easily. Indeed, the electronic configuration of $Au$ $6s^{1}$ with unpaired electron qualitatively identical with alkali metals, while electronic structure of $Zn$ and $Pb$ with paired electrons, $4s^2$ and $6p^2$ correspondingly, make them close to monoatomic molecular fluids. This is indirect support for choosing the values of $Z_c$ for lithium and sodium from \cite{book_vargaftik1975} because of similarity of electronic structure for alkali metals.

From the results described obtained above we can conclude that water, ammonia and alkali liquid metals fall into
the same class of the ILM with $q=2$ which reflects their dimerization near the critical point. Methanol is known for the
abundance of the tetramers in saturated vapor \cite{dimers_methanol_jpc1971} and identified with
$q=4$. Thus the conclusion is that low values of $Z_{c}$
for the associative LJ fluids can be attributed to the
association in near critical region. This provides the
alternative to the empirical correction factors like the Pitzer's and the Riedel's parameters at least for the fluids which demonstrate the association in near critical region. Additionally we can suggest that $Au$ is dimerized in near critical region
while $Zn$ and $Pb$ remain monoatomic. This is the difference between monovalent and divalent metals \cite{liqmetals_hensel_royalsoc1998}.

\section{Strong and weak isomorphism and the critical point line}\label{sec_weakstrict}
Simple form of the transformation \eqref{projtransfr_my} is due to the rectilinear diameter law. In particular, this means
that the CP of a given fluid $(n_c,T_c)$ lies on the binodal diameter - the median of the ``liquid-gas triangle``:
\begin{equation}\label{eq_cp_median}
  2\,\f{n_c}{n_Z}+\f{T_c}{T_Z} = 1\,.
\end{equation}
\begin{figure}
  \includegraphics[scale=1]{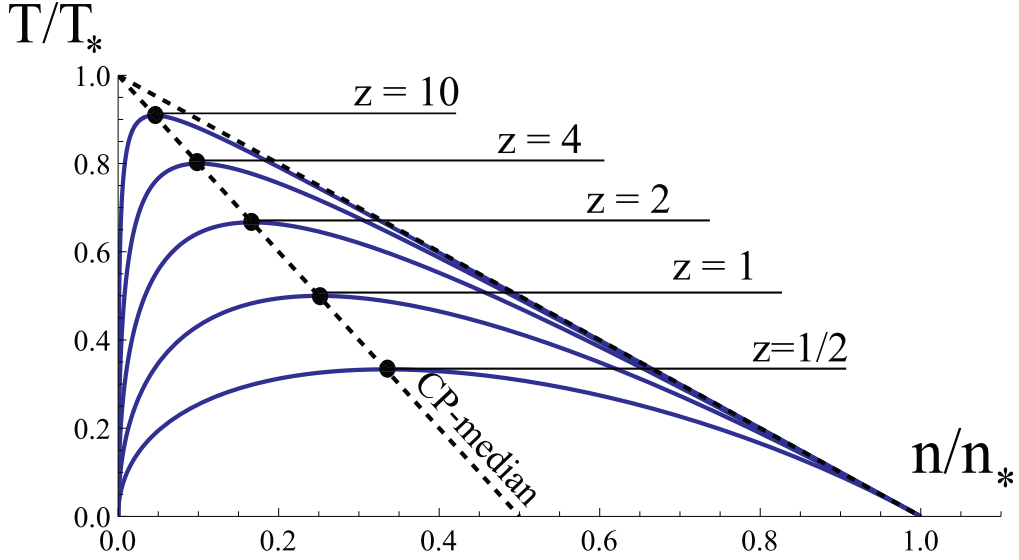}\\
  \caption{The binodals of the ILM for different values of $z$.}\label{fig_zenomedian}
\end{figure}

Also interesting empirical relation for the critical points of real fluids was revealed
in \cite{eos_zenoapfelbaum_jpchemb2008} (see also \cite{eos_zenoapfvorob_ufj2011}). It states that in reduced coordinates $n_c/n_{Z}$, $T/T_{Z}$ the critical points of the substances lie along the straight line:
\begin{equation}\label{eq_cp_vorb}
  \f{n_{c}}{n_{Z}} + \f{T_{c}}{T_{Z}} = \f{2}{3}\,.
\end{equation}
which is parallel to the Zeno line \eqref{z1}. Within the global isomorphism approach the scaling arguments in favor of the relation \eqref{eq_cp_vorb} were given  in \cite{eos_zenome_jphyschemb2010}. Here we extend these arguments in view of the results of the previous section. Indeed, using \eqref{nctcmy} it is easy to obtain:
\begin{equation}\label{eq_cp_medianmy}
  2\,\f{n_{c}}{n_{*}} + \f{T_{c}}{T_{*}} = 1\,,
\end{equation}
and
\begin{equation}\label{eq_cp_my}
  \f{n_{c}}{n_{*}} + \f{T_{c}}{T_{*}} = \f{1+2\,z}{2(1+z)}\,.
\end{equation}
For 3D LJ fluid case $z=1/2$ and Eq.~\eqref{eq_cp_my}  leads to Eq.~\eqref{eq_cp_vorb}.

The natural question arises how to explain that CPs of different fluids belong to these two lines.
To explain this we note that
\eqref{eq_cp_medianmy} does not depend on the parameter
$z = \f{T_{c}}{T_{*}-T_{c}}$. It is natural because of the Isink-like ``particle-hole`` symmetry of the binodal \cite{crit_globalisome_jcp2010}.

The global isomorphism approach states
that there are two main characteristics of the
isomorphic lattice model: the potential and the
lattice structure, e.g. elementary cell etc.
In fact there are two ways of ``deformation``. The first one is to change the attractive part of the potential thus changing the parameter $z$ (see Fig.~\ref{fig_zenomedian}) and keeping the lattice structure of the ILM fixed. In general this changes the class of the critical behavior \cite{crit_fisherlongshort_prl1972}.
Another way is to change the lattice structure with the fixed attractive potential. The class of the critical behavior does not change, only nonuniversal quantities vary.
Thus there are two levels of isomorphism:
the strong and the weak ones. The systems are isomorphic in weak
sense if their critical behavior falls in the same isomorphism class \cite{book_patpokr}.
In this sense the alkali metals, water and other LJ fluids
are (weakly) isomorphic.
For the interaction of the LJ the lattice interaction
can be taken in the simplest form of the nearest neighbor interaction. It is natural that critical asymptotics
do not depend on the fine details of the lattice structure
of the isomorphic lattice model.

\begin{figure}
\includegraphics[scale=0.75]{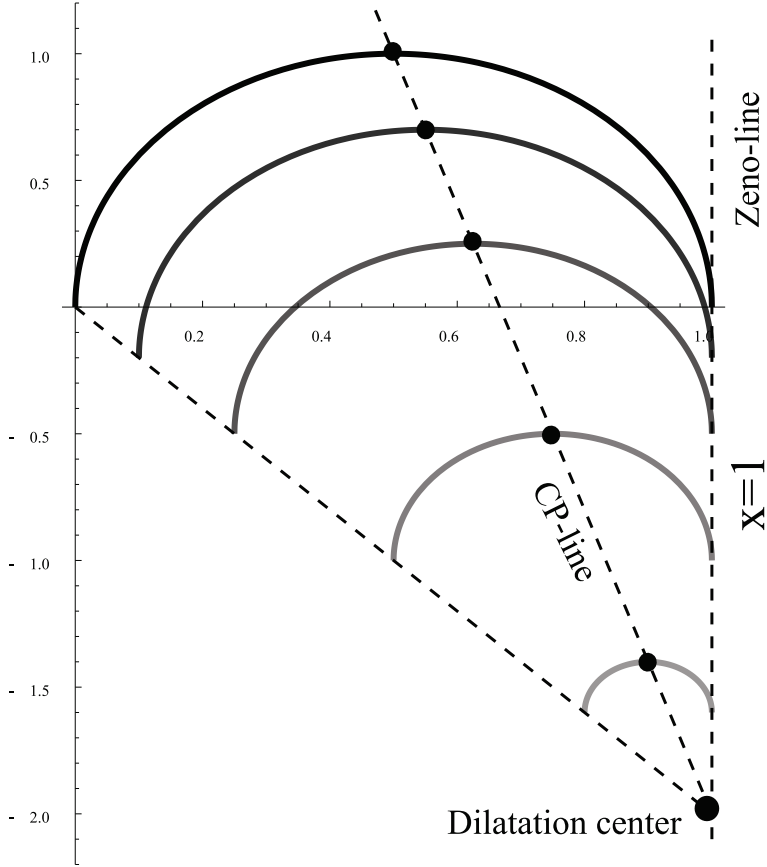}\\
  \caption{The CP line for isomorphic lattice models for $z=1/2$ (see text for the explanation).}\label{fig_cpshift}
\end{figure}

The isomorphism in the strong sense assumes
the identity of the corresponding lattice models.
Thus two LJ fluids which differ by the ILM structure
are not isomorphic in strong sense and differs by the
nonuniversal scale invariant characteristics like the critical value of the compressibility factor $Z_{c}$, etc.
Symmetric $x-t$ diagrams of these fluids can not be compared directly. Rather their symmetric ILM binodals should be related via simplest scaling relation with the Zeno-line $x=1$ as the invariant element. The result of Section~\ref{sec_zc}
assumes that the ``association`` factor $q$ introduced above plays the role of the scaling parameter. Indeed, it is directly related with the scaling of the order parameter
$x$ of the ILM. Correspondingly the temperature variable $t$ should be scaled in the proper way too. Of course, the binodals of the Ising model on different lattices are different, but their scaling properties under the scaling transformations of the basic constants - the lattice spacing and the interaction amplitude $J$ should be the same. The projective character of the transformation \eqref{projtransfr_my} allow to find easily
such scaling property which is hidden in original fluid $n-T$ diagram. Indeed, the point $x=1, \, t = -1/z$ is the center of dilation which obviously conserves the symmetry of
the binodal and represent scaling of the variables $t$ and $x$ (see Fig.~\ref{fig_cpshift}). The line joining the center of dilation with the CP
$x_c=1/2 ,\, t_c = 1$ is:
\begin{equation}\label{lgcp_line}
  \f{z}{1+z}\,t+2\,x  = \f{1+2z}{1+z}\,,
\end{equation}
and represents the line of the critical points for the lattice models generated by the corresponding scaling transformation.
It is easy to check that by the projective mapping \eqref{projtransfr_my} the line \eqref{lgcp_line}
transforms into the line of the critical points on $(n,T)$ plane Eq.~\eqref{eq_cp_my}.

In case of 3D LJ fluid when $z=1/2$ Eq.~\eqref{eq_cp_my}
perfectly corresponds numerically with the
equation of the critical point line
obtained in \cite{eos_zenoapfelbaum_jpchemb2008}
via empirical analysis of the data.
Yet from physical point of view Eq.~\eqref{eq_cp_my} differs from the original form of Apfelbaum and Vorob'ev in \cite{eos_zenoapfelbaum1_jpcb2009}. The authors of \cite{eos_zenoapfelbaum1_jpcb2009} claimed
that the right hand side depends on the critical exponent $\beta$:
\begin{equation}\label{s1}
  \f{n_{c}}{n_{Z}} + \f{T_{c}}{T_{Z}} = S_{1}(\beta)\,.
\end{equation}
But according to the definition the value $S_{1}$ depends on the locus of the critical point rather than the universal exponents. Within the concept of universality and scaling the CP locus is determined by the fluctuations of all scales while the values of critical exponents are determined by the long range fluctuations and their interactions. In terms of the lattice models this is obvious that one can change drastically CP locus by varying
the lattice properties (coordination number, symmetry etc.) without change of the universality class since $\beta$ is the invariant in a weak sense.
The same argument applies to another similarity factor:
\begin{equation}\label{s2}
S_{2} = \f{n_{c}\,T_{c}}{n_{Z}\,T_{Z}}\,(1-Z_{c})\,,
\end{equation}
which according to the results of \cite{eos_zenoapfelbaum1_jpcb2009} for the LJ fluids takes the value $S_{2}\approx 0.076$. Our results in Section~\ref{sec_zc} allow to get the estimates for $S_1,S_2$ easily in terms of our Zeno-element parameters:
\begin{equation}\label{eq_s2my}
  S_1^{*} = \f{1+2z}{2(1+z)}\,,\quad S^{*}_2 = \f{n_{c}\,T_{c}}{n_{*}\,T_{*}}\,(1-Z_{c}) = \f{z}{2(1+z)^2}\left(1-Z_c\right)\,,
\end{equation}
which for simple 3D LJ-fluid ($z=1/2,\, Z_c = 0.28$) gives:
\begin{equation}\label{eq_s12my}
S^{*}_1 = 2/3\approx 0.67\,,\quad  S^{*}_2 \approx 0.08\,\,.
\end{equation}
This perfectly matches the values obtained in \cite{eos_zenoapfelbaum1_jpcb2009} for simple LJ fluids.

So we can conclude that within the global isomorphism approach  the invariants $S_1$ and $S_2$ depend on the specific characteristics of the ILM and are invariants in the strong sense.

Note that in view of the results \cite{crit_globalisome_jcp2010} Eq.~\eqref{eq_cp_my} can be applied not only for 3D but also for 2D LJ fluids where $z=1/3$ and the critical point line is:
\begin{equation}\label{eq_cp_my2d}
  \f{n_{c}}{n_{*}} + \f{T_{c}}{T_{*}} = \f{5}{8}\,.
\end{equation}
It would be interesting to check this prediction in experiment.

\section*{Discussion}
The main result of the paper is the relation
Eq.~\eqref{zmy_c} between
critical compressibility factor of the LJ fluid and
the isomorphic lattice model which takes into account
the association of the fluid. It is based on the relation
Eq.~\eqref{zmy} with the trivial factor accounting for the association of the fluid. It supports the idea of global isomorphism relation between the equilibrium thermodynamics of the LJ fluid and the corresponding
lattice model which restores the ``particle-hole`` symmetry. Such modification reproduces known values of $Z_c$ for simple and complex LJ fluids quite good without use of any fitting parameter (see Table~\ref{tab_zclatt}). From the results we can conclude that low values of $Z_c$ signify that big fraction of associates (dimers, trimers etc) contributes to the density in near critical region. Considering the obtained results we see that apart from the liquid alkali metals mercury falls out from this class. Indeed, mercury is known for strong dependence of the interaction on the thermodynamic state \cite{liqmetals_henselhg_advphys1995}. The value of $Z_c$ for the mercury is much bigger $Z_c\approx 0.38$ and highly nonlinear
dependence of the binodal diameter takes place \cite{liqmetals_diamliqmetals_hensel_jphys1996}. At the same time mercury binodal is more symmetric in comparison with that of alkali metals. Therefore the general scheme of symmetrization can be applied \cite{eos_zenomeunified_jphyschemb2011,eos_zenovorobsymm_cpl2014}. Nevertheless, the results of \cite{liqmetals_apfelbaumvorobevhg_cml2005} suggest that for mercury the general scheme of the global isomorphism approach can be applied with the specific lattice model different from those of ``simple`` liquids. Note that in such an approach the fact that all these fluids belong to the Ising universality class of the critical behavior \cite{liqmetals_hensel_jpcondmat1990,crit_coulombfisher_jsp1994}
is due the short ranged character of the interaction. The specific features of the lattice structure affect only non universal quantities like the locus of the critical point, the critical compressibility factor $Z_c$, amplitudes etc.  This opens the way for the prediction of the high temperature critical points of metals like $Al$, $Cu$, $Fe$ based on new formulation of the thermodynamic similarity laws \cite{liqmetals_apfelbaumvorobevhg_cml2005,liqmetals_zenoapfelbaum2_cpl2009}.

Clearly the arguments presented here are of heuristic nature, yet the whole approach is based on general empirical facts and obvious symmetry properties. The problem of derivation of presented results from the first principles of statistical physics is the natural way for further studies.

\section*{Acknowledgement}
The author thank Konstantin Yun for the support of the research. The anonymous referee is acknowledged for paying attention to paper \cite{liqmetals_schroer_hthp2014}.

\newpage
%

\end{document}